\documentclass[final,authoryear,5p]{elsarticle}

\usepackage{xcolor}
\newcommand{\red}[1]{\textcolor{red}{#1}}

\usepackage{graphics} 
\usepackage{caption}
\usepackage{amssymb}
\usepackage{gensymb}

\PassOptionsToPackage{hyphens}{url} 
\usepackage[
a4paper=true,%
breaklinks=true,%
colorlinks=true,%
pdfauthor={First Author et al.},%
pdftitle={Template for manuscripts in Advances in Space Research}%
]{hyperref}

\newdimen\origiwspc
\newdimen\origiwstr

\journal{Advances in Space Research}

\begin{document}

\begin{frontmatter}

\title{MinXSS-2 CubeSat mission overview: Improvements from\\
the successful MinXSS-1 mission}

\author[1]{James Paul Mason\corref{cor}\fnref{altemail}}
\address[1]{NASA Goddard Space Flight Center, 8800 Greenbelt Rd., Greenbelt, MD, 20740, USA}
\cortext[cor]{Corresponding author}
\fntext[altemail]{\,Alternate email address: jmason86@gmail.com}
\ead{james.p.mason@nasa.gov}

\author[2]{Thomas N. Woods}
\ead{tom.woods@lasp.colorado.edu}
\author[2]{Phillip C. Chamberlin}
\ead{phil.chamberlin@lasp.colorado.edu}
\author[2]{Andrew Jones}
\ead{andrew.jones@lasp.colorado.edu}
\author[2]{Rick Kohnert}
\ead{rick.kohnert@lasp.colorado.edu}
\author[2]{Bennet Schwab}
\ead{bennet.schwab@colorado.edu}
\author[2]{Robert Sewell}
\ead{robert.sewell@lasp.colorado.edu}
\address[2]{Laboratory for Atmospheric and Space Physics, University of Colorado at Boulder, 3665 Discovery Dr., Boulder, CO, 80303, USA}

\author[3]{Amir Caspi}
\address[3]{Southwest Research Institute, 1050 Walnut St, Boulder, CO, 80302, USA}
\ead{amir@boulder.swri.edu}

\author[4]{Christopher S. Moore}
\address[4]{Harvard-Smithsonian Center for Astrophysics, 60 Garden St, Cambridge, MA, 02138, USA}
\ead{christopher.s.moore@cfa.harvard.edu}

\author[5]{Scott Palo}
\address[5]{Ann and H.J. Smead Aerospace Engineering Sciences, University of Colorado at Boulder, 429 UCB, Boulder, CO, 80303, USA}
\ead{scott.palo@colorado.edu}

\author[6]{Stanley C. Solomon}
\address[6]{National Center for Atmospheric Research, 1850 Table Mesa Dr, Boulder, CO, 80305, USA}
\ead{stans@ucar.edu}

\author[7]{Harry Warren}
\address[7]{Naval Research Laboratory, 4555 Overlook Ave SW, Washington, DC, 20375, USA}
\ead{harry.warren@nrl.navy.mil}

\begin{abstract}
The second Miniature X-ray Solar Spectrometer (MinXSS-2) CubeSat, which begins its flight in late 2018, builds on the success of MinXSS-1, which flew from 2016-05-16 to 2017-05-06. The science instrument is more advanced -- now capable of greater dynamic range with higher energy resolution. More data will be captured on the ground than was possible with MinXSS-1 thanks to a sun-synchronous, polar orbit and technical improvements to both the spacecraft and the ground network. Additionally, a new open-source beacon decoder for amateur radio operators is available that can automatically forward any captured MinXSS data to the operations and science team. While MinXSS-1 was only able to downlink about 1 MB of data per day corresponding to a data capture rate of about~1\%, MinXSS-2 will increase that by at least a factor of~6. This increase of data capture rate in combination with the mission's longer orbital lifetime will be used to address new science questions focused on how coronal soft X-rays vary over solar cycle timescales and what impact those variations have on the earth's upper atmosphere. 
\\
\\

\end{abstract}


\end{frontmatter}

\parindent=0.5 cm

\section{Introduction}
CubeSats are a valuable component of a healthy portfolio of satellite missions. While their scope is necessarily more targeted, they offer several advantages compared to large spacecraft programs. For instance, the development cycle is more brief and risks are more acceptable due to the lower budgets, which in turn allows for the development and use of truly cutting edge technologies. Additionally, constellations are only possible by scaling down the resource allocation of each individual spacecraft, which enables science that can be done no other way. Most relevantly here, they are highly compatible with the fly/learn/re-fly mentality suggested by the \cite{NRC2016}. Even though the first flight of the Miniature X-ray Solar Spectrometer \citep[MinXSS-1;][]{MasonMinXSS2016} far surpassed all success criteria, the team learned much from the spacecraft's year on orbit and those lessons have been applied to the MinXSS-2 program.

In brief, MinXSS is a 3U CubeSat with three remote-sensing instruments onboard to observe light from the sun. The primary instrument, a modified Amptek X123 silicon drift detector, measures the solar soft x-ray spectrum from 0.5 to 30~keV with nominal 0.137~keV full-width half-maximum spectral resolution at 5.9~keV \citep{Moore2016, Moore2018}. It communicates with the ground in the ultra high frequency (UHF) band. It contains a 3-axis active pointing control system provided by Blue Canyon Technologies (BCT). MinXSS-1 was the first flight of the BCT system and it was fully successful \citep{Mason2017}. MinXSS-1 was deployed from the International Space Station on 2016-05-16 at an altitude of 412~km. The planned 3-month mission extended to 12~months as the spacecraft continued to function even into the process of burning up in the atmosphere. The last received beacon was recorded at 2017-05-06 02:37:26~UTC in Australia by amateur radio operator VK2FAK. The beacon data indicated onboard temperatures upwards of 141~{\degree}C and that one of the deployable solar panels had torn off. Despite the extreme conditions of atmospheric burn up, the spacecraft was still operating. In fact, the  thermally isolated batteries were still at room temperature. More details about the thermal design, analysis, and on-orbit comparisons can be found in \citet{MasonThermal2017}. 

This paper focuses on the improvements made from MinXSS-1 to MinXSS-2. The goal of nearly all of the improvements is to receive more data on the ground. MinXSS-1 generated much more data than could fit in the narrow pipe to the ground. Both MinXSS spacecraft generate about 1~KB/s of data\footnote{\,\,housekeeping: 84.6~bytes/s; pointing telemetry: 856~bytes/s; science measurements: 25.4~bytes/s; and asynchronous log messages that are 64~bytes each} -- or 86.4~MB/day -- while in nominal science operations. MinXSS-1 only got about 15~minutes of time to downlink per day at 1200~bytes/s, so about 1~MB of downlinked data per day. Thus, we were only able to obtain about 1/86 of the data generated onboard. With the improvements made to the MinXSS-2 spacecraft, ground network, and mission profile we expect to increase the data capture percentage by at least a factor of~6 and extend the mission life by a factor of~5. In Section~\ref{sec:overview} we briefly describe the MinXSS-2 mission profile and the new science questions that can be addressed. Section~\ref{sec:improvements} describes the improvements in the spacecraft and program. Our open-source beacon decoder/forwarder is described in Section~\ref{sec:decoder} followed by a brief discussion in Section~\ref{sec:discussion}.

\section{Mission Overview}
\label{sec:overview}

\begin{figure}[bt]
\begin{center}
\includegraphics[width=8.75cm]{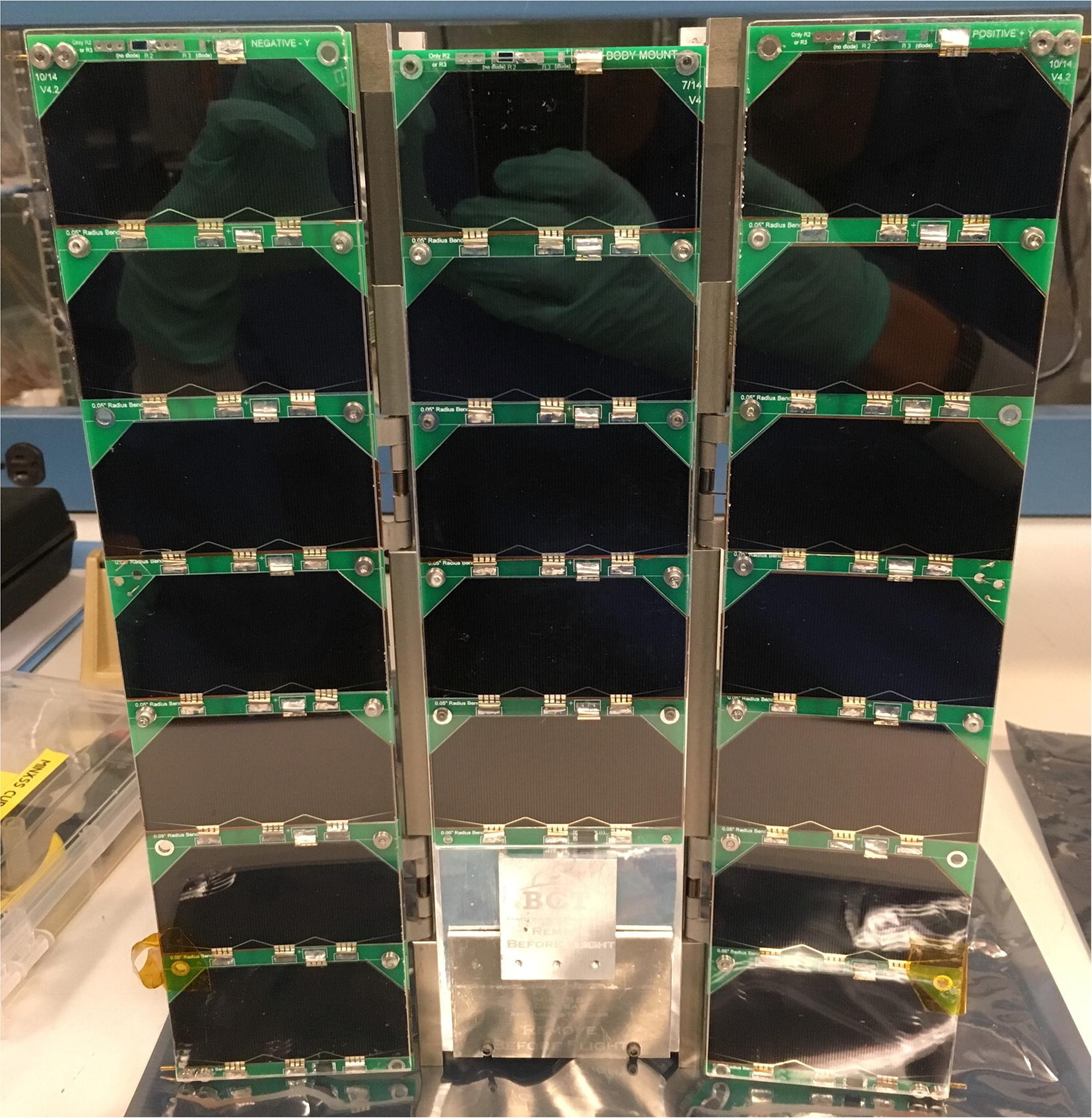}
\end{center}
\vspace{-15px}
\caption{MinXSS-2 spacecraft.}
\label{fig:minxss2}
\end{figure}

\begin{figure*}[!hbt]
\centerline{\includegraphics[width=16cm, height=10.93cm]{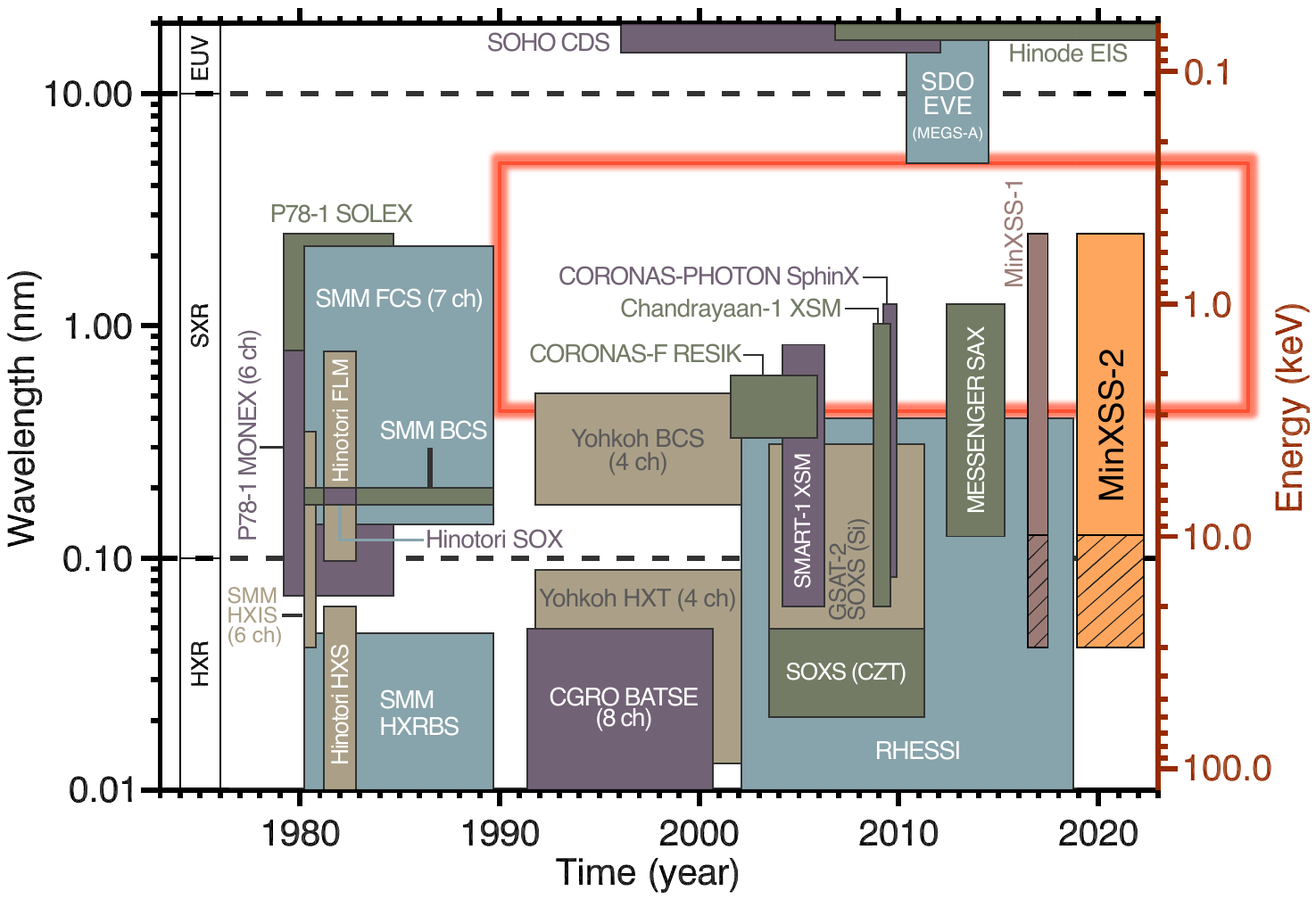}}
\vspace{-5px}
\caption{The history of solar soft X-ray measurements (excluding low-resolution broad passband instruments).}
\label{fig:sxr_history}
\end{figure*}

On 2018 December 3, MinXSS-2 (Fig.~\ref{fig:minxss2}) launched on a SpaceX Falcon~9 as part of the SSO-A ride-share mission to sun-synchronous low earth orbit. Table~\ref{tab:orbit} provides more details of the orbit. At this altitude, MinXSS-2 is expected to have significantly less atmospheric drag and thus a longer mission life. The goal is a 5-year mission. With orbital lifetime extended, the next most likely limiting factors on mission lifetime are the long term operation of the battery and the reaction wheels. The Colorado Student Space Weather Experiment \citep[CSSWE;][]{Gerhardt2013} CubeSat, which used the same type of batteries as MinXSS, operated for 2.5~years before the batteries failed. Battery lifetime is sensitive to temperature and depth of discharge, so the MinXSS-1 and -2 design runs the batteries in more optimal conditions -- shallower depth of discharge and a more stable moderate temperature -- than CSSWE did, which should extend the life. MinXSS-2 also includes flexible flight software configurations that should allow its continued science operations even with a failed battery. Specifically, once light falls on the solar arrays, the system boots up and can auto-promote itself all the way to science mode based on tuneable voltage thresholds. Reaction wheels have been a source of concern in three high-profile missions in recent years: the Kepler Space Telescope, the Hubble Space Telescope, and the Chandra X-ray Observatory \citep{Cowen2013, Dunn2018a, Dunn2018b}. BCT has been running their reaction wheels continuously on the ground for more than four years to date in order to estimate their lifetime and they have not yet failed. With no redundant wheels in MinXSS, a failure of any one of them will force the mission into a limited operational mode but it is likely that some science operations could continue. Whatever the ultimate failure mode turns out to be, we expect to obtain several years of solar soft X-ray (SXR) spectra to add to the year of observations already obtained by MinXSS-1.

\begin{table}[b]
\caption{Comparison of MinXSS-1 and -2 orbital parameters}
\label{tab:orbit}
\begin{tabular}{rll}
\hline
Parameter&MinXSS-1&MinXSS-2\\
\hline
Initial altitude & 412~km & 575~km \\
Inclination & 51.6\degree & 97.75\degree \\
Local time of descending node & various & 10:30 AM \\
$\beta$ angle $\mu$ & 32\degree & $-30$\degree \\
$\beta$ angle $\sigma$ & 19\degree & 2\degree \\
\hline
\end{tabular}
\end{table}

The majority of solar SXR observations have either been high spectral resolution over a narrow passband thus emphasizing specific plasma diagnostics, or spectrally integrated over a large passband, hence diminishing sensitivity to nuanced plasma variations (Fig.~\ref{fig:sxr_history}). MinXSS provides unique spectrally resolved solar SXR measurements over a broad passband, which maintains various plasma diagnostic capabilities without significantly compromising sensitivity to variations. The new observations from MinXSS-2 will provide a sufficiently long time baseline to study changes over solar-cycle time scales. The sun has an $\sim$11-year activity cycle that is expected to peak in 2025 \citep{Pesnell2018}, so MinXSS-2 will see the majority of the rise of the activity cycle. This will enable the science community to address new questions about the solar corona and the earth's atmospheric response. How do SXR flare energetics change with active regions of varying size and magnetic complexity over the solar cycle? What are the differences in Earth's ionosphere and thermosphere response to the solar SXR radiation across the solar cycle? Do coronal heating processes, plasma temperature, and composition change for active regions as the solar cycle evolves? By combining MinXSS observations with atomic databases and model spectra -- e.g., CHIANTI \citep{Dere1997, Landi2006} -- it is possible to estimate the composition of the corona as a function of time, which can in turn be used as discriminating evidence among competing models of coronal heating. This question can be further addressed by deriving differential emission measures (DEMs) that combine data from multiple instruments \citep{Caspi2014}. DEMs provide a description of the emitting plasma in terms of density and temperature. The approach to address the question concerning the solar-terrestrial  interaction is to include the measured solar SXR spectra as input to the National Center for Atmospheric Research (NCAR) TIME-GCM model. Historically, the results of such models have shown marked improvements when the resolution and accuracy of the solar spectral input have been increased, for example in the extreme ultraviolet \citep{Peterson2009}.

\section{Improvements Over MinXSS-1}
\label{sec:improvements}
Several improvements have been made to the MinXSS-2 mission, most of which result in higher data return. These improvements span from changes within the spacecraft and its orbit to the communications ground network. 

The earliest improvement made was obtaining an updated X123 silicon drift detector from Amptek that has better spectral resolution than the previous version and the option to increase dynamic range. In practical terms, this means that distinct emission lines will be easier to characterize in the spectra and that MinXSS-2 will be able to observe larger solar flares without the detector response becoming nonlinear. For more details see \citet{Moore2018}. Both MinXSS-2 X-ray instruments have spectral responsivity up to 30~keV, but have relatively small apertures that limit their absolute sensitivity (effective area). The magnitude of the solar photon flux exponentially decreases as photon energy increases. Thus, even for large solar flares (X-class) the measured signal will be mostly below 10--15~keV). Hence, the primary science is between 0.5--10~keV.

The orbit for MinXSS-2 also works out to be more advantageous for data return in two ways. The first was discussed in Section~\ref{sec:overview}: a higher altitude orbit results in a longer mission lifetime. That longer life means more observations will be taken overall and the dynamics of the solar cycle can be measured. The second orbital consideration is the inclination. MinXSS-2 is in a sun-synchronous, polar orbit while MinXSS-1 was in an orbit below the International Space Station at a 52{\degree} inclination. MinXSS-2's highly inclined orbit means that it will pass over ground stations worldwide and those at high latitudes tend to get about twice as many passes as those at low latitudes. To take advantage of this, the MinXSS team built a second ground station to complement the existing one. The existing ground station is at the Laboratory for Atmospheric and Space Physics (LASP) in Boulder, Colorado. 

\begin{figure}[htb]
\begin{center}
\includegraphics[width=8.75cm]{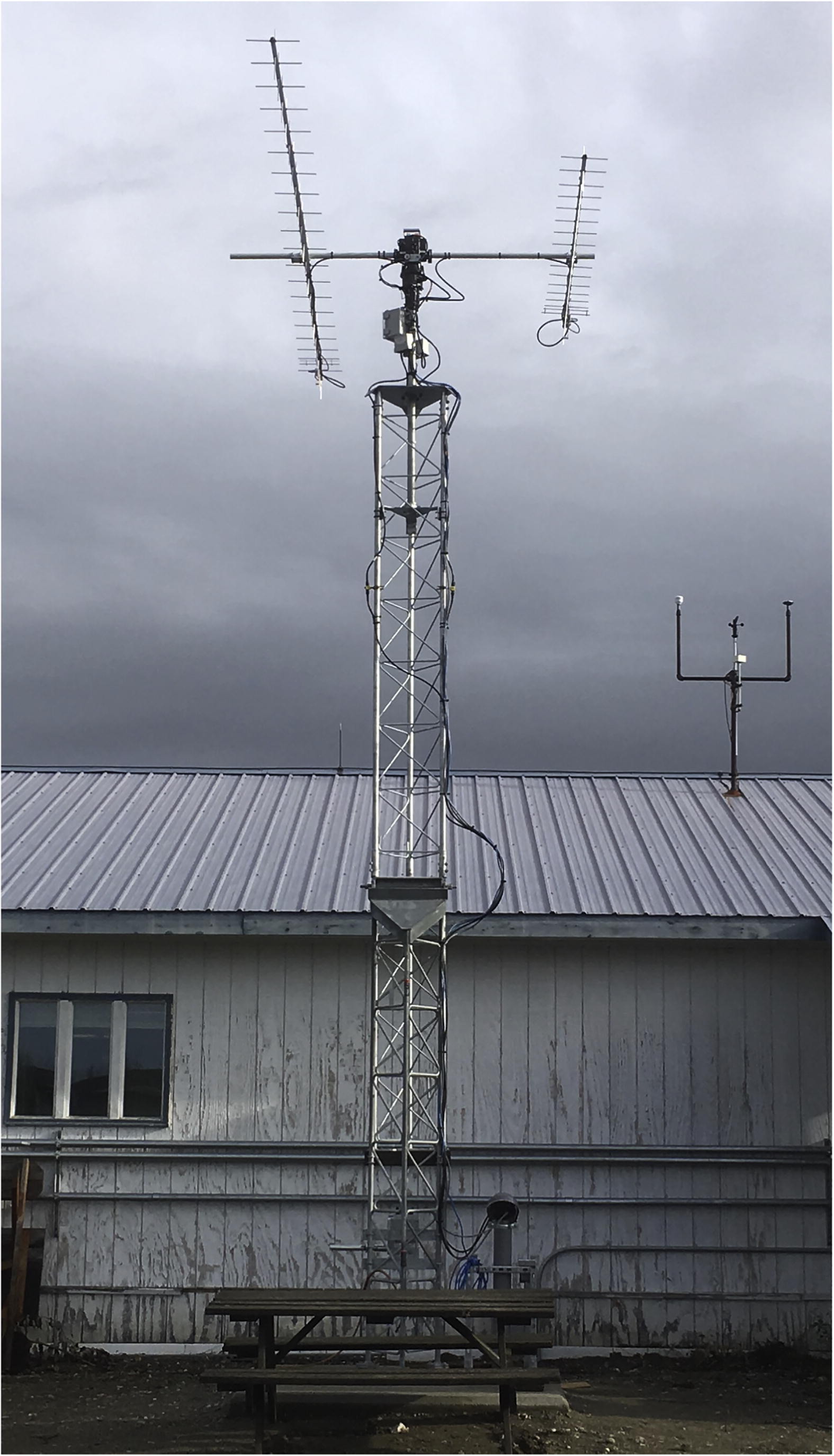}
\end{center}
\vspace{-15px}
\caption{\origiwspc=\fontdimen2=\fontdimen3\font
\fontdimen2\font=0.9\origiwspc
\fontdimen3\font=1.0\origiwstr
Ultra high frequency Yagi antennas at Alaska Satellite Facility.
\fontdimen2\font=\origiwspc
\fontdimen3\font=\origiwstr}
\label{fig:uaf_antenna}
\end{figure}

The new high-latitude ground station (Fig.~\ref{fig:uaf_antenna}) is based at the University of Alaska, Fairbanks (UAF) Geophysical Institute Alaska Satellite Facility (ASF). The average contact time\footnote{\,\,Note that these contact times include a minimum elevation mask of 10{\degree}.} at this location (6.28~min) is about the same as it is at LASP (6.46~min). Because the two ground stations are sufficiently separated, the spacecraft's field of view does not encompass both simultaneously, effectively doubling the amount of contact time the spacecraft has to downlink data. But about twice as many passes occur over ASF as at LASP, so the total data downlink time actually expands by a factor of 3: instead of $\sim$6 passes per day with only the LASP ground station, we get those 6 plus an additional 12 per day from UAF for a total of 18 passes per day. 

The next improvement is in the downlink rate of the radio. The default baud rate of MinXSS-1's AstroDev Lithium radio is 9600~baud. MinXSS-2 flight software includes a table parameter flag that allows the operators to double that rate to 19200~baud. Note that the MinXSS-1 radio had this option as well but the team decided not to implement the option for several reasons. First, the higher baud rate lowers the link margin. Because MinXSS-1 was only our second flight of a Lithium radio, we were still gaining confidence in our implementation of it and the ground station. Second, the Kenwood TS-2000 radio that we use in our ground station is not capable of supporting baud rates greater than 9600. To support 19200~baud, we needed to implement a software defined radio (SDR) solution on the ground and the team had limited experience with such systems. While SDRs provide significantly more flexibility, they also introduce a great deal of additional complexity. The team decided to err on the side of caution. Now that we have built up more confidence and have had time to implement and test a ground-based SDR, we have included the flight option to communicate at the higher baud rate. At this time of transition, the LASP and UAF ground stations have both the Kenwood radio and SDR configured in parallel with the ability to switch between them quickly and easily. As we build more confidence in our custom SDR software, we will likely leave MinXSS-2 in the 19200~baud mode. Note that the doubling of the baud rate and the higher altitude of MinXSS-2 result in decrease of about 6~dB in link margin. We make up for that by increasing the radio transmit output power by 6~dB. This higher baud rate will double our data capture rate. Combining this factor of 2 with the 3-fold increase from the UAF ground station, the total data capture rate increases by a factor of 6. 

\begin{figure}[!b]
\begin{center}
\includegraphics[width=8.75cm]{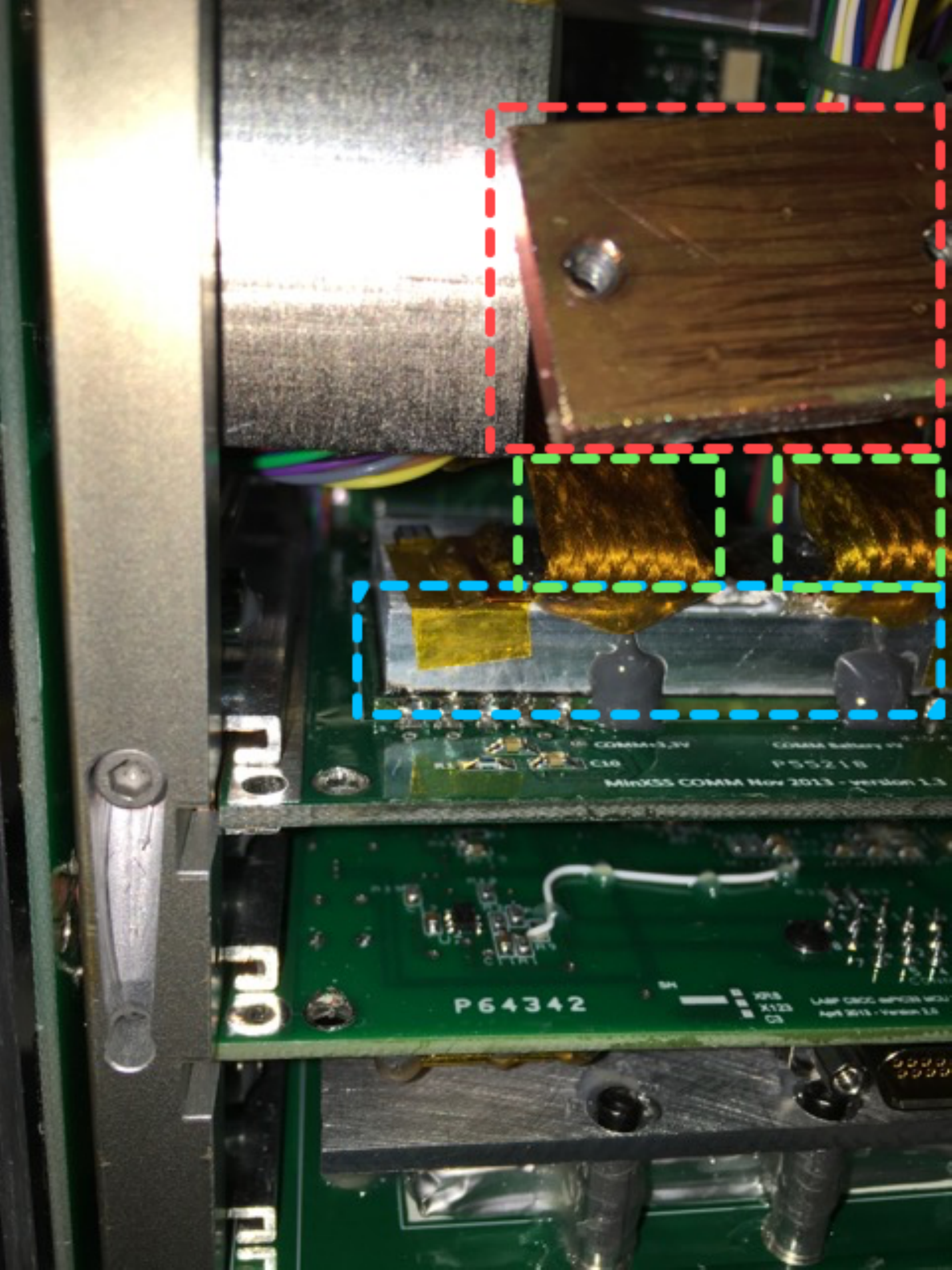}
\end{center}
\vspace{-10px}
\caption{Heat strap on MinXSS-2 radio. The red box is the copper block. The green boxes indicate the Kapton-tape-encased, braided-copper straps. The blue box indicates the radio itself.}
\label{fig:heatstrap}
\end{figure}

Another radio-related improvement was made to\linebreak MinXSS-2, this time to the hardware. On MinXSS-1 we discovered that prolonged downlink transmissions caused the radio to greatly self-heat. The temperature increase could be as much as 30~{\degree}C in a $<$10-min pass. While this increase never resulted in the radio exceeding its operational temperature limit, the downlink throughput would still sometimes become intermittent. The squelches on the Kenwood radio sounded different and packets would no longer decode. We did not keep detailed records of the number of packets lost in this manner when passes were staffed and there was no way to automate this with the Kenwood radio and terminal node controller. To avoid this issue on MinXSS-2, a heat strap was added to the radio (Fig.~\ref{fig:heatstrap}). The copper braid runs from the top of the radio housing to a copper block bolted onto the anti-sun facing side of the spacecraft. The thermal design is such that this is also effectively a dedicated radiator for the radio. The other electronics are thermally connected to two of the other faces of the spacecraft. All faces of the spacecraft have silver-coated Teflon tape to improve their emissivity and reflectivity, making them more effective radiators. In thermal vacuum testing, we found that the heat strap is highly effective: $\sim$10-min long downlinks now only raise the radio temperature by $\sim$3~{\degree}C, a 10$\times$ improvement. It is difficult to estimate how many packets may have been lost without the heat strap because we did not track the statistics, but we are confident that this improvement will mitigate that issue.

The final major improvement made to MinXSS-2 was in flight software. We now have the capability to store commands onboard along with a designated time for execution. The primary purpose of this improvement is to allow a single ground station -- the one at LASP -- to load downlink commands that will execute at specified times. With ground-based orbit prediction software, we can determine what ground stations MinXSS-2 will be in view of as a function of time. Thus, we can downlink to ground stations that we have partnerships with anywhere in the world -- including at our own second ground station at UAF -- without the need for that ground station to uplink any commands. Our baseline mission does not require this feature but it provides an advantage and a capability that future LASP CubeSat programs can leverage. This feature may increase data capture percentage beyond the 6$\times$ increase discussed above, but the value will depend on usage. Additionally, MinXSS-2 is configured to beacon a real-time housekeeping packet, nominally every 9~s, just as MinXSS-1 did. Most of these beacons are not captured by anyone but with the addition of the new MinXSS beacon decoder/forwarder software for ham radio operators, we will capture many more of these beacons.

\section{Open-Source MinXSS UHF Beacon Decoder}
\label{sec:decoder}

\begin{figure*}[ht!]
\begin{center}
\centerline{\includegraphics[width=19.3cm, height=9.3cm]{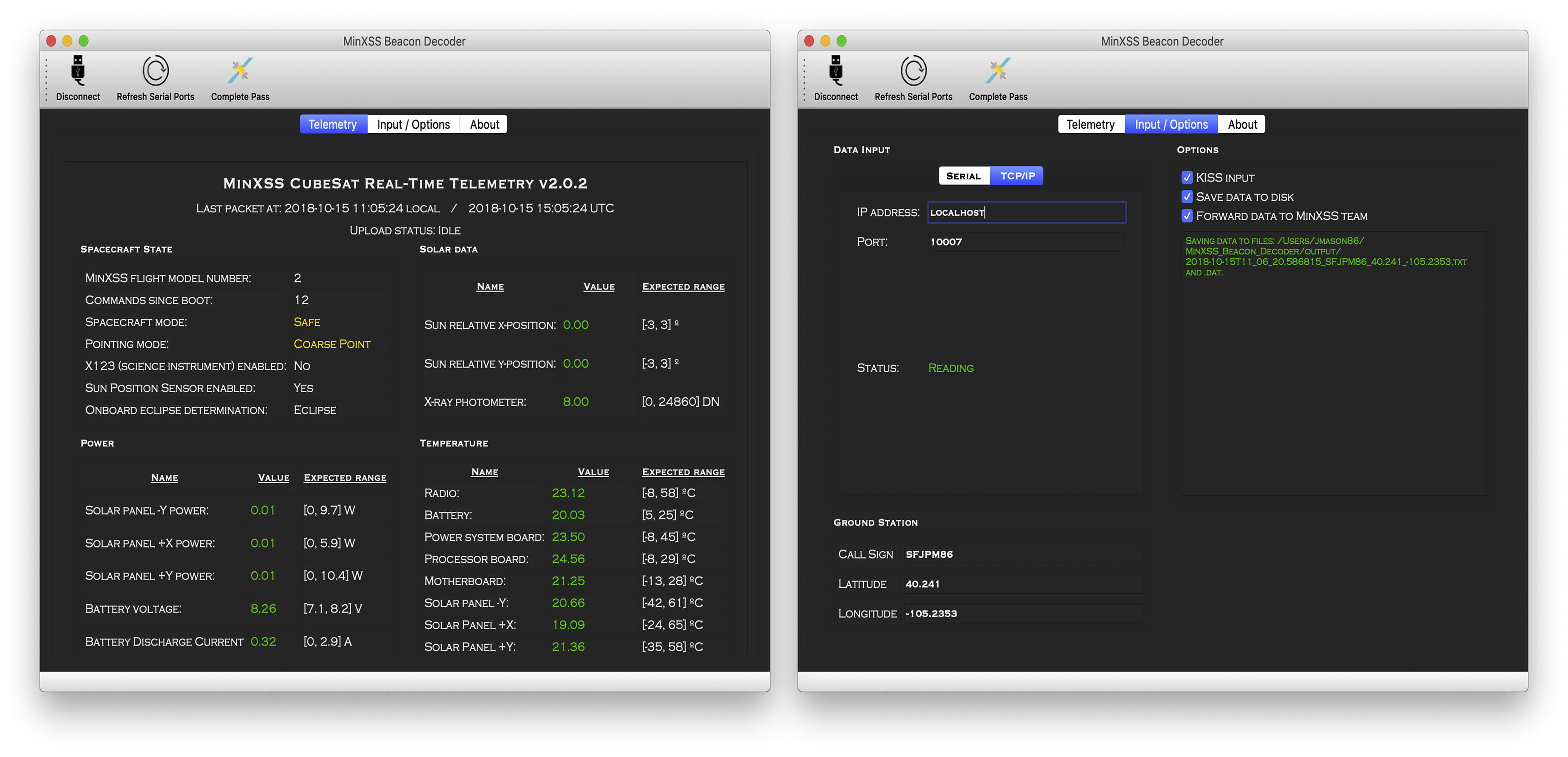}}
\end{center}
\vspace{-30px}
\caption{MinXSS ham beacon decoder/forwarder. (left) Main window; (right) input/options tab.}
\label{fig:ham}
\end{figure*}

The MinXSS beacon decoder/forwarder (Fig.~\ref{fig:ham}) has a few purposes. First, it results in more data capture for the mission. Ham operators around the world collect the MinXSS real-time beacons from their own ground stations and the application automatically forwards that data to the MinXSS team. The second purpose is to decrease the time for the operations team to be alerted to spacecraft emergencies. The network of ham operators is entirely voluntary and so there are no guarantees but any time gained is a boon. Finally, the app garners community engagement with the MinXSS program. The ham community is very active on Twitter and engages regularly with \href{https://twitter.com/minxsscubesat}{@minxsscubesat}.

The app provides an incentive for ham operators to use it: they get to see the real-time health of a spacecraft and directly contribute to a mission by capturing and forwarding its data. They also get credit for doing so on a public website where we plot the number of beacons received from each ham operator callsign. We also show a world map with the locations of all captured beacons. 

The main screen of the app displays the housekeeping telemetry that has been received and decoded. It also color codes most of the telemetry: green indicates that the telemetry point is within the expected range, yellow that it is nearing the limits, and red that the point is out of bounds or off nominal. The expected ranges are also listed explicitly. The telemetry includes spacecraft configuration, select science measurements, power, and temperatures. The second tab of the app is for configuration. The user can select whether they are using a serial line or TCP/IP for input; the former if they are working with a hardware radio and terminal node controller, or the latter if they are using a software defined radio. The user can also toggle KISS decoding, whether data save to disk, and whether the data get forwarded to the MinXSS team. The forwarding is enabled by default, and requires that the data are saved to disk. Finally, the user can optionally input identification information including their callsign and the location of their ground station. This information must be provided to be given credit on our public website but it is not necessary for data to be decoded or forwarded. All forwarded data are included in MinXSS released data products. The app is cross-platform and has been tested on macOS, Windows, and Linux. All of the software is open-source python available on GitHub \citep{MasonBeaconDecoder2018}. It was written with the hope that other CubeSat programs might fork it for their own use. 

All data received are automatically processed daily, whether they come from our own ground network, partner ground stations, or ham operators. The resultant data products are automatically released as soon as processing completes each day and are available on our public website\footnote{\,\,\url{http://lasp.colorado.edu/home/minxss/data/}}. The team is also developing real-time alerts for all incoming data. These alerts consist of emails and text messages to the operations team that are triggered any time telemetry arrives that is out of limits.

We also note that another beacon decoder/forwarder option exists\footnote{\,\,\url{http://www.dk3wn.info/software.shtml}}, written by Mike Rupprecht (callsign:\linebreak DK3WN). This Windows-only application is already widespread among ham operators. A major advantage compared to forking the MinXSS code is that it alleviates the burden on the CubeSat team to develop anything; all they need do is provide Rupprecht with their telemetry definition. The data-forwarding aspect is handled by the SatNOGS database\footnote{\,\,\url{https://db.satnogs.org/satellite/99922/}}.

The ham radio community is an excellent resource to tap into because they are worldwide, enthusiastic, and have the UHF equipment that is already compatible with the downlink capabilities of most CubeSats. As mentioned in the introduction, the final beacon from MinXSS-1 was received by a ham operator in Australia -- data that we would never have seen otherwise and that indicated the spacecraft was still operating even as it was actively burning up in the atmosphere. This was an exciting result for everyone and has become a staple of our public lectures. 

\section{Discussion}
\label{sec:discussion}

The MinXSS program has applied the build/learn/build and fly/learn/re-fly paradigms. Even with successful heritage, there are always improvements that can be made. MinXSS-1 built on the success of CSSWE, both of which were developed as student-based projects in the University of Colorado, Boulder Aerospace Engineering Sciences graduate projects course with significant involvement from professionals and use of facilities at the university's Laboratory for Atmospheric and Space Physics (LASP). LASP has been developing space-based science instruments and missions for 70 years, and still lessons are learned with every build and with every flight. MinXSS-2 extends the successful MinXSS-1 mission and will produce significantly more data due to a longer time on orbit, good alignment with our newly built second ground station, increased downlink baud rate, heat-sinked flight radio, and stored commands that allow us to downlink to any partner ground stations. While it may have been possible to upgrade MinXSS-2 to use an S- or X-band transmitter, the amount of data generated onboard doesn't justify the additional development effort it would require. The science data from MinXSS-2 are also improved with a new primary science instrument capable of greater dynamic range with higher energy resolution. Despite the highly constrained physical resources of a CubeSat, it is perfectly possible to do valuable science. 

Other CubeSat programs should also note that the relatively slow data rates inherent in UHF are not a barrier to entry even for scientifically valuable missions. We encourage them to apply the same improvements made to MinXSS-2 and to engage with the amateur radio community. With these additional data, it becomes possible to address even more scientific questions. That is precisely the plan for MinXSS-2.

\section*{Acknowledgements}
The authors would like to thank the entire MinXSS team and the wider CubeSat community. This work was supported by NASA grant NNX17AI71G. 

\bibliography{library}{}
\bibliographystyle{elsarticle-harv}

\end{document}